\documentclass[12pt]{article}

\usepackage[left=2.8cm,right=2.8cm,top=2.8cm,bottom=2.8cm]{geometry}

\usepackage[colorlinks=true,linkcolor=black,citecolor=black,urlcolor=blue,filecolor=black]{hyperref}

\usepackage{tensor}

\usepackage{color}
\usepackage{ulem}

\usepackage{amsmath}
\usepackage{amsfonts}
\usepackage{amssymb,amsthm}
\usepackage{youngtab}
\usepackage[nosort]{cite}

\usepackage{enumerate}

\numberwithin{equation}{section}

\def\d{\delta}

\renewcommand{\d}{\partial}

\newcommand{\ffrac}[2]{\raisebox{.5pt}%
  {\footnotesize$\displaystyle\frac{#1}{#2}$}\kern1pt}

\def\cI{{\cal I}}

\def\cL{{\cal L}}

\def\cR{{\cal R}}

\def\be{\begin{equation}}
\def\ee{\end{equation}}
\def\bea{\begin{eqnarray}}
\def\eea{\end{eqnarray}}
\def\ba{\begin{array}}
\def\ea{\end{array}}

\def\nn{\nonumber}

\def\12{\frac{1}{2}}

\begin{document}


\vspace{30pt}

\begin{flushright}
LPTENS-17/30
\end{flushright}

\vspace{10pt}

\begin{center}


{\Large\sc A note on ``gaugings" in four spacetime dimensions and electric-magnetic duality} 


\vspace{25pt}
{\sc Marc Henneaux${}^{\, a, b}$, Bernard Julia${}^{\, b}$, Victor Lekeu${}^{\, a}$ and Arash Ranjbar${}^{\, a}$}

\vspace{10pt}
{${}^a$\sl\small
Universit\'e Libre de Bruxelles and International Solvay Institutes, ULB-Campus Plaine CP231, B-1050 Brussels, Belgium
\vspace{10pt}

${}^b$\sl \small Laboratoire de Physique th\'eorique de l'Ecole Normale Sup\'erieure, 24 rue Lhomond, \\
75231 Paris CEDEX, France\\
\vspace{10pt}
}

\vspace{30pt} {\sc\large Abstract} 
\end{center}
 
The variety of consistent ``gauging" deformations of supergravity theories in four dimensions depends on the choice of Lagrangian formulation. One important goal is to get the most general deformations without making hidden assumptions. Ignoring supersymmetry we consider in this paper $n_v$ abelian vector potentials in four spacetime dimensions with non-minimal kinetic coupling to $n_s$ uncharged (possibly nonlinear) scalar fields. As in the case of extended supergravities, one model may possess different formulations related by $Sp(2n_v,\mathbb{R})$. The symplectic group mixes its electric and magnetic potentials. The model admits a global duality symmetry subgroup $G$ which acts also on the scalars. We recall first how the general second order Lagrangian, its local deformations and those of its abelian gauge group will depend on the choice of $2n_v$ directions (choice of ``Darboux frame"). We start from a general frame defined by the symplectic transformation relating it to a fixed``reference" one. Combinations of symplectic matrix coefficients appear then as constant parameters in the second order Lagrangians. Another gauging method uses an ``embedding tensor" that characterizes the realization of the gauge group via the global duality group. It involves additional 2-form gauge fields. A suitable zero charge limit of this realization has abelian gauge group and the ``gauging" can be viewed as a consistent deformation of that limit. We show that the two methods applied to the corresponding ungauged models have equivalent local deformations -- and more generally, have isomorphic local BRST cohomology at all ghost numbers. We finally consider manifestly duality invariant first order actions with abelian gauge group. We point out that obstructions to non-abelian deformations of the Yang-Mills type exhibited in a previous work remain present when couplings to scalar fields are included.
 

\newpage


\tableofcontents


\section{Introduction}\label{sec:intro}

It has been realized since the early days of supersymmetry that electric-magnetic duality plays a central role in four dimensional extended supergravity models \cite{Ferrara:1976iq,Cremmer:1977zt,Cremmer:1977tc,Cremmer:1977tt,Cremmer:1978ds,Cremmer:1979up,Gaillard:1981rj}.  Electric-magnetic duality is particularly relevant in the context of  so-called  gaugings of supergravities in their second order formulation since the available gaugings depend on the 
Darboux frame in which the theory is formulated. We shall follow the tradition and call this frame also the ``duality frame" or the ``symplectic frame". The last name reflects the fact that the real symplectic group preserves the symplectic tensor, written in Darboux form, on the $2n_v$ electric and magnetic field directions. We should distinguish the action of the full symplectic group acting on the frame from the ``duality" symmetry subgroup G that acts also on the scalar fields and may be strictly smaller.  

To be specific, consider  maximal supergravity \cite{Cremmer:1978km} toroidally reduced to 4 spacetime dimensions, described by the second order Lagrangian of \cite{Cremmer:1979up}.  The Lagrangian is invariant under the non-compact rigid symmetry $E_{7,7}$, which contains standard electric-magnetic duality rotations among the electric and magnetic fields.  The invariance of the equations of motion was demonstrated in \cite{Cremmer:1979up}. The invariance of the action itself is more easily verified in the first-order formalism.  This was explicitly achieved in \cite{Hillmann:2009zf}, following the crucial pioneering work of \cite{Deser:1976iy} further developed in  \cite{Henneaux:1988gg,Deser:1997mz}.

Now, while the full $E_{7,7}$ group is always present as an off-shell symmetry group of the Lagrangian, only a subgroup of it acts locally on all fields in the second-order formulation.  The other transformations act non-locally (in space).  The explicit non-local form of the duality transformations for standard electromagnetism may be found in \cite{Deser:1981fr,Bunster:2011aw}. 

The subgroup of  symmetries that acts locally on the fields of the second order formalism depends on the  symplectic frame and is called the ``electric symmetry subgroup" (in that symplectic frame).   A subgroup of this electric symmetry subgroup is the seed for the local deformations of the abelian gauge symmetry and of the action of the ``ungauged" theory usually considered in the literature.
For instance, the original gaugings of maximal supergravity were performed in \cite{deWit:1981sst} starting from the Lagrangian formulation of \cite{Cremmer:1979up} with a specific choice of 
28 ``electric" vector potentials out of the 56 potentials that one may introduce to render dualities manifest. The electric subgroup is $SL(8, \mathbb{R})$ and the subgroup into which the abelian gauge symmetry is deformed is $SO(8)$.  But changing the duality frame opens new inequivalent possibilities \cite{Andrianopoli:2002mf,Hull:2002cv}, of which \cite{DallAgata:2012mfj} provides one of the most recent examples\footnote{A systematic investigation of the space of inequivalent deformations of the Yang-Mills type has been undertaken in \cite{DallAgata:2014tph,Inverso:2015viq}.}.

Due to the dependence of the available deformations on the duality frame, it is of interest to keep the duality frame unspecified in the gaugings in order to be able to contemplate simultaneously all possibilities.  There are at least three different ways to do so.  One of them is simply to try to deform directly the manifestly duality-invariant action of \cite{Hillmann:2009zf}.  In that first-order formulation, all duality transformations are local and no choice of duality frame is needed.  However, generalizing the argument of \cite{Bunster:2010wv} to include scalar fields, we show that 
all 
standard non-abelian Yang-Mills deformations are obstructed for this Lagrangian, so that one needs to turn to  a second-order Lagrangian to have access to  non-abelian deformations   
of the Yang-Mills type. As the change of variables from the second-order formulation to the first-order one is non-local (in space), the concept of local deformations changes. 

The transition from a Hamiltonian formulation to the usual second-order formulation involves a choice of ``q"'s (to be kept) and of ``p"'s (to be eliminated)  i.e., a choice of symplectic frame.    In our case, since the electric and magnetic potentials of the first-order formulation play the respective roles of coordinates and momenta, this amounts to choosing which potentials are ``electric" and which ones are ``magnetic".  It is for that reason that this step breaks manifest duality.  One can nevertheless keep track of the symplectic transformation that relates the chosen symplectic frame to a fixed ``reference" symplectic frame.  The symplectic matrix coefficients appear then as constant parameters that define the second-order Lagrangian.   So, one is really dealing with a family of Lagrangians depending on constant parameters.   In order to treat all symplectic frames on the same footing, one should study the gaugings in a uniform manner independently of the values of these constants -- although the number and features of the solutions will depend on their actual values.   There are moreover redundancies, because many choices of the symplectic matrix coefficients, i.e.  of symplectic frame, lead to   locally equivalent second order formulations. 

A third way to keep track of the symplectic frame is through the ``embedding tensor" formalism \cite{deWit:2002vt,deWit:2005ub,deWit:2007kvg,Samtleben:2008pe,Trigiante:2016mnt},  where both ``electric" and ``magnetic" vector potentials are introduced from the beginning without choosing a priori which one is electric and which one is magnetic.  Extra $2$-forms are added in order to prevent doubling of the number of physical degrees of freedom. In addition to these fields, the Lagrangian contains also constants which form the components of the ``embedding tensor" and which indicate how the subgroup to be gauged is embedded in the duality group $G$.  These constants are not to be varied in the action -- so that  one has a second family of Lagrangians depending on external constants (with some equivalences)  -- and any choice destroys the manifest symmetry between electricity and magnetism. However if their numerical values are changed under duality transformations the latter are, 
like  B\"{a}cklund transformations, symmetries of the family of actions. 

The last two formulations are classically equivalent since the corresponding second-order Lagrangians yield equivalent equations of motion \cite{deWit:2002vt,deWit:2005ub,deWit:2007kvg,Samtleben:2008pe,Trigiante:2016mnt}.  One may, however, wonder whether the spaces of available local deformations are isomorphic.   One central result proved in this article is that this is so.  Along the way,  we clarify in detail various formulations of duality.  

At this point, a note on the terminology is perhaps not inappropriate.  The original ``ungauged"  theory is really already a gauge theory, since it has abelian vector fields, each  one with its own $U(1)$ gauge invariance.  The terminology ``gauging" might for that reason be a bit disturbing, since one is actually not gauging an initial theory without gauge invariance, but deforming the abelian gauge transformations of an original theory that is already gauged. In the process, one may charge not only  the vector fields but also the matter fields through minimal coupling.  The terminology ``charging" would be thus more accurate, but we shall follow common usage and adopt the terminology ``gauging".   We shall also use the more general terminology ``deforming", which refers to arbitrary (consistent) deformations -- and not necessarily those of the Yang-Mills type in which the potentials are viewed as connection on a 
fiber bundle, in which one ``charges" the fields as just described.  While one knows that deformations of massless vector fields are severely constrained \cite{Deser:1963zzc,Barnich:1993pa}, it is important, in a systematic investigation of the consistent interactions, to make no a priori assumption.

Our paper is organized as follows.
\begin{itemize}
 \item In Section \ref{sec:symplectic}, we recall the properties of the first-order, manifestly duality invariant action.   We stress in the first subsection how the symplectic group emerges in the discussion of electric-magnetic duality in four spacetime dimensions, and more generally, contrast the cases of $D= 4m$ and $D= 4m+2$ spacetime dimensions.  Returning to $4$ dimensions  we then also show that the obstructions to local non-abelian deformations of the Yang-Mills type
exhibited in  \cite{Bunster:2010wv}, remain valid when couplings to scalar fields are included.  This result, which was actually already sketched in \cite{Bunster:2011aw}, forces one to go to 
second-order formulations in order to allow the 
local gaugings.
 \item Next, in Section \ref{sec:duality} we discuss in depth this transition to the usual second-order formalism, which involves a choice of symplectic frame.   We indicate how to keep explicit track of that choice in the formalism through the matrix of the symplectic transformation associated with the chosen symplectic frame.  We point out that in fact, what really matters for the transition to the second-order formalism, is not the symplectic frame itself, but the equivalence class of symplectic frames defining the same Lagrangian submanifold for the ``q"'s.  We define the terms ``symplectic = Darboux = duality" frame and ``electric (stability) group".  We also  distinguish local changes of variables from the non-local steps  that imply a change of class of local deformations.
 
 \item We then show in Section \ref{sec:embedding} how to reformulate the gaugings of the embedding tensor method as deformations of an appropriate zero-charge limit and observe that each embedding tensor model  \cite{deWit:2002vt,deWit:2005ub,deWit:2007kvg,Samtleben:2008pe,Trigiante:2016mnt} 
can indeed be recovered as a local deformation of this abelianized limit. We demonstrate that the latter's space of local deformations is isomorphic to the space of local deformations of the standard second-order action with the appropriate choice of the corresponding  duality frame. This is because one of the conditions fulfilled by the embedding tensor of  \cite{deWit:2002vt,deWit:2005ub,deWit:2007kvg,Samtleben:2008pe,Trigiante:2016mnt}, namely their ``locality" constraint, implies precisely that there exists a frame in which all the gauge couplings are electric.   By expressing the theory in that frame and eliminating the redundant (auxiliary or pure gauge) fields, one goes to the standard formulation. This elimination procedure does not change the space of local deformations so that  both formulations have exactly the same local BRST cohomology  at ghost number zero - the relevant cohomology for the deformations \cite{Barnich:1993vg}. 

In fact, as we show in this paper, all local cohomology groups $H^{k}(s \vert d)$ are isomorphic in both formulations. In particular the potential 
(or ``candidate") 
anomalies (cohomology at ghost number one) are also the same. Accordingly, one can analyze the consistent deformations in either formulation.  In a companion paper written in collaboration with G. Barnich and N. Boulanger,  the consistent deformations are systematically explored in the standard second-order formulation \cite{preparation}.
 \item We close our paper with conclusions (Section \ref{sec:conclusions}).
\end{itemize}

Since the main features related to dualities and duality frames are already present 
in simpler four-dimensional models consisting of a collection of $n_v$ abelian vector fields $ A_\mu^I$
coupled to $n_s$ scalar fields $\phi^i $ in such a way that there is a 
non-trivial duality group $G \subset Sp(2n_v)$, we only consider those models in the present paper. 
More specifically, the ``zero charge", second order Lagrangians are\footnote{Our conventions are as follows: the spacetime metric is the Minkowski metric with signature $(-,+,\cdots, +)$.  The Levi-Civita tensors $\varepsilon^{\lambda \mu \nu \rho}$ in $4$ spacetime dimensions and $\varepsilon^{ijk}$ in $3$ spatial dimensions are such that $\varepsilon^{0123} = -1$ (so that $\varepsilon_{0123} = 1$) and $\varepsilon^{123} = 1$. }
\be
\cL^{(2)} = \cL^{(2)}_S + \cL_V^{(2)}, 
\ee
 where
\begin{align} 
\cL^{(2)}_V &= - \frac{1}{4} \mathcal{I}_{IJ}(\phi) F^I_{\mu\nu} F^{J\mu\nu} + \frac{1}{8} \mathcal{R}_{IJ}(\phi)\, \varepsilon^{\mu\nu\rho\sigma} F^I_{\mu\nu} F^J_{\rho\sigma} \label{eq:lag}
\end{align}
with $ F^I_{\mu\nu} = \partial_\mu A^I_\nu -  \partial_\nu A^I_\mu $, and the scalar Lagrangian is taken to be 
\be
\cL^{(2)}_S = -\frac{1}{2} g_{ij}(\phi) \d_\mu \phi^i \d^\mu \phi^j,
\ee
but its explicit form is actually not crucial in our derivations.
Neglecting gravity, this is the generic bosonic sector of ungauged supergravity. The matrices $\cI$ and $\cR$ give the non-minimal couplings between the scalars and the abelian vectors. 
The generating set of gauge invariances is given before ``gauging" by
\begin{equation}
\label{eq:gauge}
\delta A_\mu^I=\d_\mu\epsilon^I,\quad \delta\phi^i =0. 
\end{equation}

We close this introduction by emphasizing that we develop the analysis in a manner that does not make any a priori assumption on the type of deformations, even when we discuss the embedding tensor formalism that was tailored for deformations of the Yang-Mills type.  The only exception is the no-go theorem of Section \ref{sec:symplectic}, which specifically forbids Yang-Mills deformations in the first-order formalism with both electric and magnetic potentials.

\section{Symplectic structure in electric-magnetic internal space}
\label{sec:symplectic}

\subsection{$4m$ versus $4m+2$ spacetime dimensions}
For a $p$-form in $D= 2p + 2$ spacetime dimensions, the electric and magnetic fields are exterior forms of same rank $p+1$ and one may consider electric-magnetic duality transformations that mix them while leaving the theory invariant.   It turns out that the cases $p$ odd ($p= 2m - 1$, $D= 4m$) and $p$ even ($p=2m$, $D=4m +2$) lead to different duality groups. The electric-magnetic duality group is a subgroup of $Sp(2n_v, \mathbb{R})$ in $4m$ spacetime dimensions and $O(n_v, n_v)$ in $4m+2$ spacetime dimensions \cite{Deser:1997mz,Deser:1997se,Julia:1997cy}.   The easiest way to see this is to go to the Hamiltonian formalism.  

We first consider the case $p=1$ of interest here ($D=4$).

The canonical momenta conjugate to the $A^I$ are given by
\begin{equation}
\pi^i_I = \frac{\d \mathcal{L}}{\d \dot{A}^I_i} = \mathcal{I}_{IJ}\, F\indices{^J_0^i} -\frac{1}{2} \mathcal{R}_{IJ}\,\varepsilon^{ijk}F^J_{jk}, 
\end{equation}
along with the constraint $\pi^0_I=0$. This relation can be inverted to get
\begin{equation}
\dot{A}^{Ii} = (\mathcal{I}^{-1})^{IJ} \,\pi^i_J + \partial^i A^I_0 + \frac{1}{2} (\mathcal{I}^{-1} \mathcal{R})\indices{^I_J} \,\varepsilon^{ijk} F^J_{jk} ,
\end{equation}
from which we can compute the first-order Hamiltonian action
\begin{equation}
S_H =  \int\!d^4x\, \left( \pi^i_I \dot{A}^I_i - \mathcal{H} -  A^I_0 \, \mathcal{G}_I \right),
\end{equation}
where
\begin{align}
\mathcal{H} &= \frac{1}{2} (\mathcal{I}^{-1})^{IJ} \pi^i_I \pi_{J i} + \frac{1}{4} (\mathcal{I} + \mathcal{R}\mathcal{I}^{-1}\mathcal{R})_{IJ} F^I_{ij} F^{Jij} + \frac{1}{2} (\mathcal{I}^{-1} \mathcal{R})\indices{^I_J} \,\varepsilon^{ijk} \pi_{Ii} F^J_{jk} \\
\mathcal{G}_I &= - \partial_i \pi^i_I .
\end{align}
The time components $A^I_0$ appears in the action as Lagrange multipliers for the constraints
\begin{equation}
\partial_i \pi^i_I = 0 .
\end{equation}
These constraints can be solved by introducing new (dual) potentials $Z_{I i}$ through the equation
\begin{equation}
\pi^i_I = -  \varepsilon^{ijk} \partial_j Z_{I k},
\end{equation}
which determines $Z_{I}$ up to a gauge transformation $Z_{I i} \rightarrow Z_{I i} + \partial_i \tilde{\epsilon}_I$. Note that the introduction of these potentials is non-local but permitted in (flat) contractible space. 
Putting this back in the action gives
\begin{equation} \label{eq:symlag}
S = \frac12 \int \!d^4x \left(  \Omega_{MN} \mathcal{B}^{Mi} \dot{\mathcal{A}}^N_i -   \mathcal{M}_{MN}(\phi) \mathcal{B}^M_i \mathcal{B}^{Ni} \right),
\end{equation}
where the doubled potentials are packed into a vector
\begin{equation}
(\mathcal{A}^M) = \begin{pmatrix} A^I \\ Z_I \end{pmatrix}, \quad M= 1, \dots, 2n_v,
\end{equation}
and their curls  $\mathcal{B}^{Mi}$ are
\begin{equation}
\mathcal{B}^{Mi} =  \varepsilon^{ijk} \partial_j \mathcal{A}^M_k .
\end{equation}
The matrices $\Omega$ and $\mathcal{M}(\phi)$ are the $2n_v \times 2n_v$ matrices
\begin{equation} \label{eq:OMdef}
\Omega = \begin{pmatrix}
0 & I \\ -I & 0
\end{pmatrix}, \qquad
\mathcal{M} = \begin{pmatrix}
\mathcal{I} + \mathcal{R}\mathcal{I}^{-1}\mathcal{R} & - \mathcal{R} \mathcal{I}^{-1} \\
- \mathcal{I}^{-1} \mathcal{R} & \mathcal{I}^{-1}
\end{pmatrix},
\end{equation}
each block being $n_v\times n_v$.  The dual vector potentials are canonically conjugate to the magnetic fields.  The action (\ref{eq:symlag}), which puts electric and magnetic potentials on the same footing, is called throughout this paper the ``first-order action".

The symmetric matrix $\mathcal{M}$ can be decomposed as $\mathcal{M}=\mathcal{S}^T \mathcal{S}$ where the symplectic matrix $\mathcal{S}$ is given by 
$\mathcal{S} =  \mathcal{E} \Delta$,  where
\begin{equation} \label{eq:OEdef} 
\mathcal{E} = \begin{pmatrix}
f & 0 \\
0 &  f^{-T}
\end{pmatrix},
\end{equation}
 with $f^{-T}\equiv (f^{-1})^T$ and
$\mathcal{I} = f^T f$
and where we defined
\begin{equation} 
\Delta =
 \begin{pmatrix}
1 & 0 \\
-\mathcal{R} & 1
\end{pmatrix}.
\end{equation}
This form is familiar in the case of homogeneous scalar manifolds  \cite{Trigiante:2016mnt} (section 2).

The kinetic term for the potentials in the first-order action (\ref{eq:symlag}) can be rewritten 
\be
\frac12\int dt \, d^3x \, d^3y \, \sigma_{M N}^{ij}(\vec{x}, \vec{y}) \,  \mathcal{A}^{M}_i(\vec{x}) \, \dot{\mathcal{A}}^{N}_j(\vec{y}), \label{eq:presymp} \ee
where $\sigma_{M N}^{ij}(\vec{x}, \vec{y})$ defines a (pre-)symplectic structure in the infinite-dimensional space of the $\mathcal{A}^M_i(\vec{x})$ and is antisymmetric for the simultaneous exchange of $(M,i,\vec{x})$ with $(N,j,\vec{y})$,
\be
\sigma_{M N}^{ij}(\vec{x}, \vec{y}) = - \sigma_{N M}^{ji}(\vec{y}, \vec{x}).
\ee
Explicitly, $\sigma_{M N}^{ij}(\vec{x}, \vec{y})$ is given by the product 
\be
\sigma_{M N}^{ij}(\vec{x}, \vec{y})=  -\Omega_{MN} \varepsilon^{ijk} \partial_k \delta(\vec{x} - \vec{y})
\ee
The internal part $\Omega_{MN}$ is antisymmetric under the exchange of $M$ with $N$, while the spatial part $\varepsilon^{ijk} \partial_k \delta(\vec{x} -\vec{y})$ is symmetric under the simultaneous exchange of $(i,\vec{x})$ with $(j,\vec{y})$.    A necessary
condition for a linear transformation in the internal space of the potentials to be a symmetry of the action is that it should leave
 the kinetic term invariant, which in turn implies that $\Omega_{MN}$ must remain invariant under these transformations, i.e., they must belong to the symplectic group $Sp(2n_v, \mathbb{R})$.

However for $2$-forms $\mathcal{A}^M_{ij}(\vec{x})=  -\mathcal{A}^M_{ji}(\vec{x})$ in six spacetime dimensions, the expression analogous to (\ref{eq:presymp}) is
 \be
\frac12\int dt \, d^5x \, d^5y \, \sigma_{M N}^{ij \, pq}(\vec{x}, \vec{y}) \,  \mathcal{A}^{M}_{ij}(\vec{x}) \, \dot{\mathcal{A}}^{N}_{pq}(\vec{y}), \label{eq:presymp2} \ee
where the (pre)symplectic form  $\sigma_{M N}^{ij \, pq}(\vec{x}, \vec{y})$ is again given by the product of a matrix in internal space and a matrix involving the spatial indices (including ($\vec{x}$)),
\be
\sigma_{M N}^{ij \, pq}(\vec{x}, \vec{y})=  -\eta_{MN} \varepsilon^{ijpqk} \partial_k \delta(\vec{x}- \vec{y}).
\ee
As a presymplectic form,  $\sigma_{M N}^{ij \, pq}(\vec{x}, \vec{y})$ must be antisymmetric for the simultaneous exchange of $(M,ij,\vec{x})$ with $(N,pq,\vec{y})$.  But now, the spatial part $\varepsilon^{ijpqk} \partial_k \delta(\vec{x} - \vec{y})$ is {\it antisymmetric},
$$ \varepsilon^{ijpqk} \partial_k \delta(\vec{x}- \vec{y}) = - \varepsilon^{pqijk} \partial_k \delta(\vec{y}- \vec{x}) $$ 
and therefore the internal part $\eta_{MN}$ must be {\it symmetric}.  And indeed, this is what one finds \cite{Henneaux:1988gg,Bekaert:1998yp}.  The signature of $\eta_{MN}$ is $(n_v,n_v)$ and its diagonalization  amounts to split the $2$-forms into chiral (self-dual) and anti-chiral (anti-self-dual) parts.   A necessary condition for a linear transformation in the internal space of the potentials  
to be a symmetry of the action is again that it should leave the kinetic term invariant, but this implies now that $\eta_{MN}$ must be invariant under these transformations, i.e., they must belong to the orthogonal group $O(n_v,n_v)$.

The analysis proceeds in exactly the same way in higher (even) dimensions.  The presymplectic form $\sigma_{M N}^{i_1 i_2 \cdots i_p \, j_1 j_2 \cdots j_p}(\vec{x}, \vec{y})$ for the $p$-form potentials $\mathcal{A}^M_{i_1i_2 \cdots i_p}(\vec{x})$ is equal to the product $-\rho_{MN} \varepsilon^{i_1 i_2 \cdots i_p  j_1 j_2 \cdots j_p k} \partial_k \delta(\vec{x}- \vec{y}) $.  The spatial part $ \varepsilon^{i_1 i_2 \cdots i_p  j_1 j_2 \cdots j_p k} \partial_k \delta(\vec{x}- \vec{y}) $ is symmetric in spacetime dimensions $D$ equal to $4m$ ($p$ odd) and antisymmetric in spacetime dimensions $D$ equal to $4m + 2$ ($p$ even).  Accordingly the matrix $\rho_{MN}$ in electric-magnetic internal space must be -- and is! -- antisymmetric for $D=4m$ and symmetric for $D=4m+2$, implying that the relevant electric-magnetic group is a subgroup of $Sp(2n_v, \mathbb{R})$ or of $O(n_v,n_v)$. 
These results were established in \cite{Deser:1997mz,Deser:1997se,Julia:1997cy} and already anticipated in \cite{Julia1981} (see also \cite{Cremmer:1997ct,Bremer:1997qb,Deser:1998vc,Julia:2005wg} for independent developments).  As stressed in \cite{Julia:1997cy}, they do not depend on the spacetime signature.

\subsection{Global duality group in $4$ dimensions}
\label{sub:DualGroup}

We now come back to the case $p=1$ in four dimensions described by the Lagrangian $\mathcal{L} = \mathcal{L}^{(2)}_S + \mathcal{L}_V^{(1)}$ where $\mathcal{L}_V^{(1)}$ is the vector Lagrangian in first order form,
\be
\mathcal{L}_V^{(1)} = \frac{1}{2} \Omega_{MN} \mathcal{B}^{Mi} \dot{\mathcal{A}}^N_i - \frac{1}{2} \mathcal{M}_{MN}(\phi) \mathcal{B}^M_i \mathcal{B}^{Ni} \label{eq:LV1}
\ee
(see (\ref{eq:symlag})).

Of special interest are the symmetries of  $\mathcal{L}$ for which the vector fields transform linearly,
\be 
\delta \mathcal{A}^M_i(\vec{x}) = \epsilon^\alpha \left(t_\alpha\right)^M_{\; \; N} \mathcal{A}^N_i(\vec{x}) \label{eq:trans1}
\ee
accompanied by a transformation of the scalars that may be nonlinear,
\be
\delta \phi^i = \epsilon^{\alpha} R_{\alpha}^{\; \; i}(\phi). \label{eq:trans2}
\ee
Here, the $\epsilon^\alpha$'s are infinitesimal parameters.  One might consider more general (nonlinear canonical) transformations of the vector fields, and the scalars might have their own independent symmetries, but we shall confine our attention in this paper to symmetries of this type as they encompass what is usually referred to as ``duality symmetries".

For the action to be invariant, the transformation (\ref{eq:trans1}) should not only be symplectic, but also such that the accompanying transformation (\ref{eq:trans2}) of the scalar field is given by a Killing vector of the scalar metric $g_{ij}(\phi)$ (so as to leave the scalar Lagrangian $\cL_S$ invariant) which implies that
\be
\epsilon^{\alpha} \left(\frac{\partial  \mathcal{M}_{MN}}{\partial \phi^i} R_{\alpha}^{\; \; i} + \mathcal{M}_{MP} \left(t_\alpha\right)^P_{\; \; N}
+ \mathcal{M}_{PN} \left(t_\alpha\right)^P_{\; \; M} \right)= 0 \label{eq:trans3}
\ee
(so as to leave the energy density of the electromagnetic fields invariant).  In finite form, the symmetries are 
$\mathcal{A} \rightarrow \bar{\mathcal{A}}$, $\phi \rightarrow \bar{\phi}$ where the symplectic transformation
\begin{equation}
\bar{\mathcal{A}} = F \mathcal{A}, \label{eq:Trans1}
\end{equation}
 and the isometry $\phi \rightarrow \bar{\phi}$ of the scalar manifold are such that
\begin{equation}
\mathcal{M}(\bar{\phi}) = F^{-T} \mathcal{M}(\phi) F^{-1} .\label{eq:Trans3}
\end{equation}

The algebra of transformations of the type (\ref{eq:trans1}), (\ref{eq:trans2}) that leave the action invariant is called the ``(electric-magnetic)  duality algebra" $\mathcal{G}$, while the corresponding group is the ``(electric-magnetic) duality group" $G$. 
As we have just indicated, the duality algebra is a subalgebra of $sp(2n_v, \mathbb{R})$.  This symplectic condition holds irrespective of the scalar sector and its couplings to the vectors, since it comes from the invariance of the vector kinetic term, which assumes always the same form.  Which transformations among those of $sp(2n_v, \mathbb{R})$ are actually symmetries depend, however, on the number of scalar fields, their internal manifold and their couplings to the vectors \cite{Ferrara:1976iq,Cremmer:1977zt,Cremmer:1977tc,Cremmer:1977tt,Gaillard:1981rj,deWit:2001pz}.  

If there is no scalar field, the invariance of the Hamiltonian $\frac{1}{2} \delta_{MN} \mathcal{B}^M_i \mathcal{B}^{Ni}$ in (\ref{eq:LV1}) implies that a duality transformation should also belong to $so(2n_v)$ so as to leave the metric $\delta_{MN}$ invariant, and so the duality algebra is $sp(2n_v, \mathbb{R}) \cap so(2n_v) \equiv u(n_v)$.  The duality group is the unitary group $U(n_v)$.  If there are scalars forming the coset manifold $Sp(2n_v, \mathbb{R})/U(n_v)$, with appropriate couplings to the vectors, the duality algebra is the full symplectic algebra $sp(2n_v, \mathbb{R})$ and the duality group is $Sp(2n_v, \mathbb{R})$.  In the case of maximal supergravity toroidally reduced to four dimensions, the scalars are such that the duality group is $E_{7,7}$ as discovered in \cite{Cremmer:1978ds,Cremmer:1979up}.

\subsection{No deformations of the Yang-Mills type for the first-order action}

It was shown in \cite{Bunster:2010wv} that the first order action (\ref{eq:LV1}) without scalar fields,
\be
 \frac12\int dt\, d^3x \left(\Omega_{MN} \mathcal{B}^{Mi} \dot{\mathcal{A}}^N_i -  \delta_{MN} \mathcal{B}^M_i \mathcal{B}^{Ni}\right) \label{eq:LV1bis}
\ee
does not admit non-abelian deformations of the Yang-Mills type. 

 That is, there is no subgroup of the duality group $U(n_v)$ that can be deformed into a non-abelian one.  A much stronger result was actually derived earlier in \cite{Bekaert:2001wa} through the BRST formalism, namely, that the action (\ref{eq:LV1bis}) admits no local deformation that deforms the gauge algebra at all.

In fact, the obstruction described in \cite{Bunster:2010wv}  does not depend on the scalar sector and obstructs Yang-Mills deformations even when scalar fields are present.  The clash comes from the incompatibility of an adjoint action (as required by the Yang-Mills construction) and the symplectic condition (as required by the invariance of the scalar-independent kinetic term). The persistence of the obstruction even in the presence of scalar fields was announced in the conclusion of \cite{Bunster:2011aw}, but the proof was only sketched there.  For completeness, we give the details here.

To be more specific, consider the Yang-Mills deformation 
\be
\delta \mathcal{A}^M_i = \partial_i \mu^M + g C^M_{\; \; NP} \mathcal{A}^N_i \mu^P 
\ee
of the original abelian gauge symmetry of (\ref{eq:LV1bis}),
\be
\delta \mathcal{A}^M_i = \partial_i \mu^M
\ee
with gauge parameters $\mu^M$.   Here, the $C^M_{\; \; NP} = - C^M_{\; \; PN}$ are the structure constants of the gauge group $G_g$ of dimension $2n_v$ into which the original abelian gauge group  is deformed, and $g$ is the deformation parameter.

To make the abelian gauge invariance of the starting point more manifest, one can introduce the temporal component $\mathcal{A}^M_0$ of the potentials and rewrite the action (\ref{eq:LV1bis}) in terms of the abelian curvatures $f_{\mu \nu}^M = \partial_\mu  \mathcal{A}^M_\nu - \partial_\nu  \mathcal{A}^M_\mu$.  One can replace $\dot{\mathcal{A}}^N_i$ by $f_{0i}^M$ in (\ref{eq:LV1bis}) because the magnetic fields are identically transverse.

Under the Yang-Mills deformation, the abelian curvatures are replaced by the non-abelian ones,
\be
F_{\mu \nu}^M = f_{\mu \nu}^M + g C^M_{\; \; NP} \mathcal{A}^N_\mu \mathcal{A}^P_\nu,
\ee
which transform in the adjoint representation as
\be
\delta F_{\mu \nu}^M = g C^M_{\; \; NP} F^N_{\mu \nu} \mu^P, \label{eq:adjoint}
\ee
and the ordinary derivatives $\partial_\mu \phi^i$ of the scalar fields are replaced by the covariant derivatives $D_\mu \phi^i$.  These contain linearly the undifferentiated vector potentials $\mathcal{A}^N_\mu$.  

The deformed action reads
\be
\frac{1}{2} \int d^4 x \, g_{ij}(\phi) D_\mu \phi^i D^\mu \phi^j + \frac14 \int d^4 x \left(\Omega_{MN} \epsilon^{ijk}F^M_{ij} F^N_{0k} - \mathcal{M}_{MN}(\phi) F^M_{ij} F^{Nij} \right)
\ee
The key observation now is that the kinetic term for the vector potentials is the same as in the absence of scalar fields.  Invariance of the kinetic term under the Yang-Mills gauge transformations  yields therefore the same conditions as when there is no $\phi^i$ (the kinetic term must be invariant by itself since the scalar Lagrangian is invariant and there can be no compensation with the energy density $\mathcal{M}_{MN}(\phi) F^M_{ij} F^{Nij}$ that contains no time derivative).  These conditions are that the adjoint representation (\ref{eq:adjoint}) of the gauge group $G_g$ should preserve the symplectic form $\Omega_{MN}$ in internal electric-magnetic space \cite{Bunster:2010wv} and read explicitly
\be
C_{NPM} = C_{MPN} \label{eq:InvSymp}
\ee
with $C_{MNP} \equiv \Omega_{MQ}C^Q_{\; \; NP}$. The symmetry of $C_{MNP}$ under the exchange of its first and last indices, and its antisymmetry in its last two indices, force it to vanish.  Hence, the structure constants $C^Q_{\; \; NP}$ must also vanish and there can be no non-abelian deformation of the gauge symmetries.  The Yang-Mills construction in which the potentials become non-abelian connections  is unavailable in the manifestly duality invariant first-order formulation \cite{Bunster:2010wv}\footnote{Other types of deformations - charging the matter fields or adding functions of the abelian curvatures - are not excluded by the argument since they preserve the abelian nature of the gauge group.}.  In order to have access to these deformations, one must avoid the condition (\ref{eq:InvSymp}) expressing the invariance of the symplectic form.  This is what is effectively achieved when one goes to the second-order formalism by choosing a Lagrangian submanifold, on which the pull-back of $\Omega_{MN}$ is by definition zero.  This is explained in the next section.

\section{Choice of symplectic frames for the transition to the second order formalism and electric group ambiguity}\label{sec:duality}

While the first-order and second-order formalisms are equivalent in terms of symmetries (any symmetry in one formalism is also a symmetry of the other\footnote{This is standard material, see for instance exercise 3.17 of \cite{Henneaux:1992ig}.}), they are not equivalent in what concerns the concept of locality (specifically, locality in space).  A local function in one formulation may not be local in the other.  The origin of this difference is made more precise.  Consequently, the spaces of consistent local deformations are not the same in both formalisms.

\subsection{Choice of symplectic frame and locality restriction}\label{sec:symplecticchoice}

In order to go from the first-order formalism to the standard second-order formalism, one has to tell what are the ``$q$"'s (to be kept) and the ``$p$"'s (to be eliminated in favour of the velocities through the inverse Legendre transformation).  
We have actually made the choice to use Darboux coordinates in \eqref{eq:OMdef} and 
we call the corresponding frame a Darboux frame or symplectic frame.

In our case, since the electric and magnetic fields are conjugate, a choice of $q$'s and $p$'s is equivalent to  choosing what are the ``electric potentials" and what are the conjugate ``magnetic potentials".   For that reason,    a choice of Darboux frame is also called a choice of ``duality frame".  

Since we consider here only linear canonical transformations, a choice of canonical coordinates amounts to a choice of an element of the symplectic group $Sp(2n_v, \mathbb{R})$ relating that symplectic frame (i.e., symplectic basis) to  a reference symplectic frame.

Once a choice of symplectic frame has been made, one goes from (\ref{eq:symlag}) to the second-order formalism by following the  steps reverse to those that led to the first-order formalism:
\begin{itemize}
\item One keeps the first half of the vector potentials $\mathcal{A}^I = A^I$, the ``electric ones" in the new frame ($I = 1, \cdots, n_v$).
\item One replaces the second half of the vector potentials $\mathcal{A}^{I +n_v}= Z_I$ (the ``magnetic ones") by the momenta
\be
\pi^i_I = - \varepsilon^{ijk} \partial_j Z_{I k}
\ee
subject to the constraints 
\begin{equation}
\partial_i \pi^i_I = 0 
\end{equation}
which one enforces by introducing the Lagrange multipliers $A_0^I$.  This is the non-local step (in space).
\item One finally eliminates the momenta $\pi^i_I$, which can be viewed as auxiliary fields, through their equations of motion which are nothing else but the inverse Legendre transformation expressing $\dot{A}^I_i$ in terms of $\pi^i_I$.
\end{itemize}

One can encode the choice of symplectic frame through the symplectic transformation that relates  the chosen symplectic frame to a  reference symplectic frame,
\begin{equation} \label{eq:Aprime}
\mathcal{A} = E \mathcal{A}',
\end{equation}
where $E$ is a symplectic matrix in the fundamental representation of $ Sp(2n_v, \mathbb{R})$,  while $\mathcal{A}$ and $\mathcal{A}'$ are respectively the potentials in the reference and new duality frames. The symplectic property $E^T \Omega E = \Omega$ ensures that the kinetic term in \eqref{eq:symlag} remains invariant. The first-order action therefore takes the same form, but with matrices $\mathcal{M}$ and $\mathcal{M}'$  related by 
\begin{equation}\label{eq:Mprime}
\mathcal{M}' = E^T \mathcal{M} E .
\end{equation}

Using the fact that $E$ and $E^T$ are symplectic, a straightforward but not very illuminating computation shows that $\mathcal{M}'$ determines matrices $\mathcal{I}'$, $\mathcal{R}'$ uniquely such that equation \eqref{eq:OMdef} with primes holds,  i.e, there exist unique $\mathcal{I}'$ and $\mathcal{R}'$ such that
\begin{equation} \label{eq:OMdefBis}
\mathcal{M}' = \begin{pmatrix}
\mathcal{I}' + \mathcal{R}'\mathcal{I}'^{-1}\mathcal{R}' & - \mathcal{R}' \mathcal{I}'^{-1} \\
- \mathcal{I}'^{-1} \mathcal{R'} & \mathcal{I}'^{-1}
\end{pmatrix}.
\end{equation}
The matrices $\mathcal{I}'$ and $\mathcal{R}'$ depend on the scalar fields, but also on the symplectic matrix $E$.
Performing in reverse order the steps described above,  one then gets the second-order Lagrangian  
\begin{equation} \label{eq:lprime}
\cL^{'2}_V = - \frac{1}{4} \mathcal{I}'_{IJ}(\phi) F'^I_{\mu\nu} F'^{J\mu\nu} + \frac{1}{8} \mathcal{R}_{IJ}'(\phi)\, \varepsilon^{\mu\nu\rho\sigma} F'^I_{\mu\nu} F'^J_{\rho\sigma} .
\end{equation}

The new Lagrangian \eqref{eq:lprime} depends on the parameters of the symplectic transformation $E$ used in equation \eqref{eq:Mprime}.  So we really have a family of Lagrangians labelled by an $Sp(2n_v, \mathbb{R})$ element $E$.  By construction, these Lagrangians differ from one another by a change of variables that is in general non-local in space. 

There are of course redundancies in this description. The presence of these redundancies will be important in the study of deformations.

First,  it is clear that different symplectic transformations can lead to the same final $n_v$-dimensional ``Lagrangian subspace" of the $q$'s upon elimination of the momenta.  [The choice of the $q$'s is equivalent to a choice of Lagrangian linear subspace, or linear polarization, because the $q$'s form  a complete set of commuting variables (in the Poisson bracket). The same is true for a choice of $p$'s.] The stability subgroup $T$ of a Lagrangian subspace consists of the block lower triangular symplectic transformations, yielding second-order Lagrangians that differ by a redefinition of the $q$'s (when $E$ is a canonical ``point" transformation) and the addition of a total derivative (when $E$ is a canonical ``phase" transformation).  Interestingly, the matrix $\mathcal{S} =  \mathcal{E} \Delta$ introduced above belongs to this stability subgroup $T$. 

Second, 
duality symmetries correspond also to redundancies in the transition to the second-order formalism since they do not modify the first-order Lagrangian (when the scalars are transformed appropriately) and hence clearly lead to the same final second-order Lagrangian.

One can characterize the redundancies as follows.  The symplectic transformations \eqref{eq:Aprime} with matrices $E \in Sp(2 n_v, \mathbb{R})$ and $g E t$, where $t$ belongs to the stability subgroup $T$ of the final Lagrangian subspace and $g$ belongs to the duality group $G$, are equivalent. Indeed, the first-order Lagrangian is left invariant under the transformation  $\mathcal{A} \rightarrow g^{-1} \mathcal{A}$ (provided the scalars are transformed appropriately), and two sets of new symplectic coordinates $\mathcal{A}'$ and $ t \mathcal{A}'$ yield equivalent second-order Lagrangians after elimination of the momenta\footnote{One could also consider what happens when the coordinates $\mathcal{A}'$ defined by \eqref{eq:Aprime} are taken as the reference frame. Then, the changes of frame $\mathcal{A}' = S \mathcal{A}''$ and $\mathcal{A}' = g' S t' \mathcal{A}''$ are equivalent, where $g' = E^{-1} g E$ is the matrix associated with the duality symmetry $g$ in the $\mathcal{A}'$--frame and $t'$ is an element of the stability subgroup of the Lagrangian subspace in the $\mathcal{A}''$--frame. This gives an equivalent description of the redundancies.}. The relevant space is thus the quotient $G \backslash Sp(2n_v, \mathbb{R})/ T$.  [In terms of symplectic bases rather than coordinates, one should use dual (transposed) actions of the groups.]  It should be noted that the authors of \cite{deWit:2002vt} considered a smaller stability subgroup $T = GL(n_v,\mathbb{R})$ by requiring  invariance of the Lagrangian itself (and not invariance up to a total derivative that does not matter classically).  

We have repeatedly emphasized that a local expression in the $Z^I_k$'s need not be local when expressed in terms of the $\pi^k_I$'s and thus of the Lagrangian variables.  This is because the magnetic potentials are not local functions of the conjugate momenta.  But the $\pi^k_I$'s are local functions of the $Z^I_k$'s given by $\pi^i_I = - \varepsilon^{ijk} \partial_j Z_{I k}$ and so it would seem that any local function of the $\pi^i_I$'s, and thus of the Lagrangian variables, should also be a local function of the $Z^I_k$'s.  
This is true as long as the relation $\pi^i_I = - \varepsilon^{ijk} \partial_j Z_{I k}$ is not modified.  In deformations of the second-order Lagrangians, one allows, however, deformations that modify this relation.  For instance,  after Yang-Mills deformations,  the abelian constraints $\partial_i\pi^i_I = 0$ get replaced by non-abelian ones and their local solution gets replaced by a non-local expression involving both kinds of potentials.

\subsection{Electric group}

The duality transformations
$\mathcal{A} \rightarrow \bar{\mathcal{A}} = F \mathcal{A}$, $\phi \rightarrow \bar{\phi}$, where the symplectic matrix $F$  and the isometry $\phi \rightarrow \bar{\phi}$ of the scalar manifold are such that the condition (\ref{eq:Trans3}) holds, 
generically mix the electric and magnetic potentials.  Accordingly, when expressed in terms of the variables of the second-order formalism, they will generically take a non-local form, since the magnetic potentials become non-local functions of the electric potentials and their time derivatives in the second-order formalism \cite{Deser:1976iy,Deser:1981fr,Bunster:2011aw}\footnote{Given a duality symmetry of the action $S[\mathcal{A}^I_k, \phi^i]$, one gets the corresponding duality symmetry of the first-order action $S[A^I_k, \pi_I^k, A_0^I]$ by (i) expressing in the variation $\delta A^I_k$ the magnetic potentials $Z^I_k$ (if they occur) in terms of $\pi_I^k$, which is a non-local expression, determined up to a gauge transformation that can be absorbed in a gauge transformation of the electric variables;  (ii) computing the variation $\delta \pi^k_I$ from  $\pi^i_I = - \varepsilon^{ijk} \partial_j Z_{I k}$; and (iii) determining the variation of the  Lagrange multipliers $A_0^I$ so that the terms proportional to the constraint terms $\partial_i \pi^i_I$ cancel in $\delta S[A^I_k, \pi_I^k, A_0^I]$.  One gets the symmetry of the second order action by expressing the auxiliary fields $\pi^i_I$ that are eliminated in terms of the retained variables.}$^,$\footnote{For comparison with \cite{Gaillard:1981rj}, we stress that the duality transformations are defined here in terms of the fundamental variables of the theory that are varied in the action principle, namely the potentials, following \cite{Deser:1976iy}.  These transformations imply on-shell the duality transformations of the field strengths considered in \cite{Gaillard:1981rj}.}.

Thus, an electric symmetry transformation is characterized by the property that the matrix $F$ is lower-triangular, i.e., given that it must be symplectic
\begin{equation} \label{eq:Ge0}
 F = \begin{pmatrix} M & 0 \\ BM & M^{-T} \end{pmatrix} , \; M \in GL(n_v), \; B^T = B.
\end{equation}
An electric symmetry is therefore a transformation $\bar{A}^I_\mu = M\indices{^I_J} A^J_\mu$ of the electric potentials for which there is a symmetric matrix $B$ and an isometry $\phi \rightarrow \bar{\phi}$ of the scalar manifold such that
\begin{equation} \label{eq:Ge}
\mathcal{M}(\bar{\phi}) = F^{-T} \mathcal{M}(\phi) F^{-1}
\end{equation}
with $F$ given by (\ref{eq:Ge0}).  The terminology ``electric group" for the group of transformations characterized by (\ref{eq:Ge0}) is standard.  The transformations with $B \not=0$ involve transformations of Peccei-Quinn type (axion shift symmetry) \cite{Peccei:1977hh,Peccei:1977ur}. 

The electric group $G_e$ in a given frame depends of course on the chosen duality frame. Indeed, going to another duality frame with the symplectic matrix $E$ as in \eqref{eq:Aprime} will replace the matrices $F$ by their conjugates $F' = E^{-1} F E$, and these might not be lower-triangular.
The condition that $F'$ has the lower-triangular form \eqref{eq:Ge0} therefore depends on the choice of $E$.
 
The interest of the electric group $G_e$ is that its transformations are local in the second-order formalism.  Therefore, among the duality transformations, it is the electric ones that are candidates for (local) gaugings. 

\section{Deformation leading to the embedding tensor formalism}\label{sec:embedding}
\subsection{Magnetic potentials, two forms and abelianization} \label{subsec:doubling}

Another way to cover arbitrary choices of symplectic frames is given by the embedding tensor formalism of \cite{deWit:2005ub,deWit:2007kvg,Samtleben:2008pe,Trigiante:2016mnt}. 
It involves a collection of extra fields. In addition to the $n_v$ ``electric" vector fields $A^I_\mu$, the Lagrangian also contains $n_v$ ``magnetic" vector fields $\tilde{A}_{I\mu}$ and $\dim(G)$ two-forms $B_{\alpha \mu \nu}$ ($\alpha = 1, \dots, \dim(G)$), where $G$ is the full duality group (in which one can also include pure scalar symmetries). The Lagrangian of \cite{deWit:2005ub} also depends on the ``embedding tensor" $\Theta\indices{_M^\alpha}=(\Theta\indices{_I^\alpha},\Theta\indices{^{I\alpha}})$, which are constants in spacetime and not to be varied in the action. They contain information about the embedding of the gauge group in the duality group through the connection 
\begin{equation}
\label{Theta}
\mathcal{T}_\mu^\alpha = \mathcal{A}_\mu^M \Theta_M^\alpha
\end{equation}
which is $\mathcal{G}$-valued (where $\mathcal{G}$ is the Lie algebra of $G$) and where we wrote $\mathcal{A}^M_{\mu} = (A^I_\mu, \tilde{A}_{I\mu})$. $\tilde{A}_{Ik}$ was called $Z_{Ik}$ above. The embedding tensor also  contains information about the choice of symplectic frame  -- a fact relevant for our purposes --  since, as shown in \cite{deWit:2005ub}, one of the constraints on the embedding tensor that is present in this approach  implies that the ``gauged" group must be a subgroup of an electric subgroup of $G$. 

The ``gauged" Lagrangian of reference \cite{deWit:2005ub} reads
\begin{align}
\mathcal{L}^{\Theta, \text{int.}}(A,\tilde{A},B) = &- \frac{1}{4} \mathcal{I}_{IJ}(\phi) \mathcal{H}^I_{\mu\nu} \mathcal{H}^{J\mu\nu} + \frac{1}{8} \mathcal{R}_{IJ}(\phi)\, \varepsilon^{\mu\nu\rho\sigma} \mathcal{H}^I_{\mu\nu} \mathcal{H}^J_{\rho\sigma} \nonumber \\
&- \frac{g}{8} \varepsilon^{\mu\nu\rho\sigma} \Theta^{I\alpha}B_{\alpha \mu\nu} \left( \tilde{\mathcal{F}}_{I \rho\sigma} - \frac{g}{4} \Theta\indices{_I^\beta} B_{\beta \rho\sigma}\right) + g \mathcal{L}^\text{extra} (A,\tilde{A}, gB ) .
\label{eq:19}
\end{align}
where
\begin{equation}
\mathcal{H}^I_{\mu\nu} = \mathcal{F}^I_{\mu\nu} + \frac{g}{2} \Theta^{I \alpha} B_{\alpha \mu\nu} 
\end{equation}
and $\mathcal{F}^I_{\mu\nu}$, $\tilde{\mathcal{F}}_{I\mu\nu}$ are the Yang-Mills curvatures, differing from the abelian ones by $O(g)$-terms.
The ``extra" terms in \eqref{eq:19} are terms necessary to secure gauge invariance but which vanish in the limit taken below. They are not written for that reason and their explicit form may be found in \cite{deWit:2005ub}. As we emphasized in \eqref{eq:19}, they depend on the two-forms $B_\alpha$ only through the combination $B{''}_\alpha = g B_\alpha$ and carry at least an extra factor of $g$. The $B{''}_\alpha$'s are necessary to shift away some field strengths even for $g=0$. The authors of \cite{deWit:2005ub} also consider matter couplings that we do not write here as they are irrelevant for our purposes.
Note that the full duality group is in general not faithfully represented.

The  gauged Lagrangian \eqref{eq:19} is an interacting Lagrangian.
To view it as a deformation of an ungauged Lagrangian with the same field content, so as to phrase the gauging problem as a deformation problem, we observe that the straightforward free limit $g\rightarrow 0$ of that Lagrangian is just the Lagrangian \eqref{eq:lag}, without any additional fields and without the embedding tensor. But if we redefine the two-forms appearing in the Lagrangian \eqref{eq:19} as $B{''}_\alpha = g B_\alpha$ and then take the limit $g \rightarrow 0$, one gets
\begin{equation} \label{lagtwoforms}
\mathcal{L}^\Theta = - \frac{1}{4} \mathcal{I}_{IJ}(\phi) H^I_{\mu\nu} H^{J\mu\nu} + \frac{1}{8} \mathcal{R}_{IJ}(\phi)\, \varepsilon^{\mu\nu\rho\sigma} H^I_{\mu\nu} H^J_{\rho\sigma} - \frac{1}{8} \varepsilon^{\mu\nu\rho\sigma} \Theta^{I\alpha}B{''}_{\alpha \mu\nu} \left( \tilde{F}_{I \rho\sigma} - \frac{1}{4} \Theta\indices{_I^\beta} B{''}_{\beta \rho\sigma}\right),
\end{equation}
with
\begin{equation}
H^I_{\mu\nu} = F^I_{\mu\nu} + \frac{1}{2} \Theta^{I \alpha} B{''}_{\alpha \mu\nu} 
\end{equation}
and $F^I_{\mu\nu}$, $\tilde{F}_{I \mu\nu}$ the abelian curvatures.  This Lagrangian contains the extra fields and has 
the ability to cover generic symplectic frames through  the embedding tensor components $\Theta^{I\alpha}$ and $\Theta\indices{_I^\alpha}$. These satisfy the  so-called ``locality" quadratic constraint
\begin{equation}
\Theta^{I [\alpha} \Theta\indices{_I^{\beta]}} = 0 . \label{eq:ThetaCons}
\end{equation}
In addition to (\ref{eq:ThetaCons}), the embedding tensor of \cite{deWit:2005ub} fulfills other constraints.  We stress that (\ref{eq:ThetaCons}) is the only constraint relevant to our analysis -- it is the only one that we will use. 
Let us note that any term vanishing because of these extra constraints could be added to the above Lagrangian, but for definiteness we adopt (\ref{lagtwoforms}).

The Lagrangian \eqref{lagtwoforms} is invariant under the $2n_v$ gauge transformations
\begin{align}
\delta A^I_\mu &= \partial_\mu \lambda^I - \frac{1}{2} \Theta^{I \alpha} \Xi{''}_{\alpha \mu} \\
\delta \tilde{A}_{I\mu} &= \partial_\mu \tilde{\lambda}_I + \frac{1}{2} \Theta\indices{_I^\alpha} \Xi{''}_{\alpha \mu}
\end{align}
and the $\dim G$ generalized gauge transformations
\begin{align}
\delta B{''}_{\alpha\mu\nu} &= 2 \partial_{[\mu}\Xi{''}_{\alpha \nu]},
\end{align}
where we redefined the gauge parameter as $\Xi{''}_\alpha = g \Xi_\alpha$ with respect to \cite{deWit:2005ub}.   The parameters $\lambda^I$, $\tilde{\lambda}_I$ and $\Xi{''}_{\alpha \nu}$ are arbitrary functions.  The $\lambda$'s  and $\tilde{\lambda}$'s correspond to standard $U(1)$ gauge symmetries of the associated one-form potentials.
The $\Xi''$'s define ordinary abelian two-form gauge symmetries and also appear as shift transformations of the one-form potentials. This set of gauge symmetries is reducible since adding a gradient to $\Xi''$ and shifting simultaneously the $\lambda^I$ and $\tilde{\lambda}_I$, i.e.\begin{equation}
\Xi{''}_{\alpha \mu} \rightarrow \Xi{''}_{\alpha \mu} + \partial_\mu \xi_{\alpha}, \qquad \lambda^I \rightarrow \lambda^I + \frac{1}{2} \Theta^{I\alpha} \xi_{\alpha}, \qquad \tilde{\lambda}_I \rightarrow \tilde{\lambda}_I - \frac{1}{2} \Theta\indices{_I^\alpha} \xi_\alpha,
\end{equation}
leads to no modification of the gauge transformations.

By construction, the non-abelian Lagrangian of \cite{deWit:2005ub} can be viewed as a consistent local deformation of (\ref{lagtwoforms}),  since one can charge the fields in (\ref{lagtwoforms}) in a smooth way -- by adding $O(g)$ and $O(g^2)$ terms -- to get the non-abelian Lagrangian.  The abelian Lagrangian (\ref{lagtwoforms}) is thus a sensible starting point for the deformation procedure.  The space of consistent local deformations of (\ref{lagtwoforms}) will necessarily include the non-abelian Lagrangian of \cite{deWit:2005ub}\footnote{If the embedding tensor components $\Theta^{I\alpha}$ and $\Theta\indices{_I^\alpha}$ do not fulfill the extra constraints of \cite{deWit:2005ub}, one will simply find that there is no deformation of the type of \cite{deWit:2005ub} among the consistent deformations of (\ref{lagtwoforms}).}. 

A natural question to be asked is whether the space of local deformations  of (\ref{lagtwoforms}) is isomorphic to the space of local deformations of the conventional Lagrangian \eqref{eq:lprime} in an appropriate symplectic frame.  We show that this is the case.  

To that end, we will now perform a sequence of field redefinitions and show that the Lagrangian (\ref{lagtwoforms}) with a given definite (arbitrary) choice of embedding tensor differs from the Lagrangian \eqref{eq:lprime} in a related symplectic frame by the presence of algebraic auxiliary fields that can be eliminated without changing the local BRST cohomology.  The other extra fields are pure gauge with gauge transformations that are pure shifts so that they do not appear in the Lagrangian and do not contribute either to the local BRST cohomology.  We recall that the local BRST cohomology at ghost number zero is what controls the space of non trivial deformations \cite{Barnich:1993vg}.

The steps performed below for $g=0$  follow closely the steps given in \cite{deWit:2005ub} to prove that the extra fields appearing in the embedding tensor formalism do not add new degrees of freedom with respect to conventional gaugings.  As shown by the examples of the first-order and second-order formulations discussed in the previous sections, however, equivalent actions might admit inequivalent spaces of local deformations.  Another example is given in \cite{Brandt:1996au,Brandt:1997uq,Brandt:1997ny}.   We  therefore feel  that it is  important to relate the local BRST cohomologies, and we monitor in particular carefully below how locality goes through each field redefinition.   Furthermore, at no stage do we fix the gauge in order to make it clear that the analysis is gauge-independent.  This is also important since gauge fixing might have an impact on local cohomology.

The magnetic components $\Theta^{I\alpha}$ of the embedding tensor form a rectangular $n_v \times (\dim G)$ matrix. Let $r \leq n_v$ be its rank.
Then, there exists an $n_v \times n_v$ invertible matrix $Y$ and a $(\dim G) \times (\dim G)$ invertible matrix $Z$ such that
\begin{equation} \label{eltheta}
\Theta^{' I \alpha} = Y\indices{^I_J} \Theta^{J \beta} Z\indices{_\beta^\alpha}
\end{equation}
is of the form
\begin{equation} \label{formthetaprime}
(\Theta^{' I \alpha}) = \left(\begin{array}{c|c}
\theta & 0  \\ \hline
0 & 0
\end{array}\right),
\end{equation}
where $\theta$ is an $r\times r$ invertible matrix. (Indeed, the matrices $Y$ and $Z$ can be constructed as a product of the matrices that implement the familiar ``elementary operations" on the rows and columns of $\Theta^{I \alpha}$.) 

We also define the new electric components by
\begin{equation} \label{magtheta}
\Theta\indices{^'_I^\alpha} = (Y^{-1})\indices{^J_I} \Theta\indices{_J^\beta} Z\indices{_\beta^\alpha},
\end{equation}
so that the new components still satisfy the constraint (\ref{eq:ThetaCons})
\begin{equation} \label{primedconstraint}
\Theta^{' I [\alpha} \Theta\indices{^'_I^{\beta]}} = 0 .
\end{equation}
As in section 5.1 of \cite{deWit:2005ub}, let us split the indices as $I = (\hat{I}, \hat{U})$ and $\alpha = (i, m)$, where $\hat{I}, i = 1, \dots, r$, $\hat{U} = r+1, \dots, n_v$ and $m = r+1, \dots, \dim G$. We also define
\begin{equation}
\tilde{\theta}\indices{_{\hat{I}}^i} = \Theta\indices{^'_{\hat{I}}^i} .
\end{equation}
With this split, equations \eqref{formthetaprime} and \eqref{primedconstraint} become
\begin{align} \label{thetasplit}
\Theta^{'\hat{I}i} &= \theta^{\hat{I}i} \quad \text{(invertible)} \nn \\
\Theta^{'\hat{I}m} &= \Theta^{'\hat{U}i} = \Theta^{'\hat{U}m} = \Theta\indices{^'_{\hat{I}}^m} = 0 \nn \\
\theta^{\hat{I}i} \tilde{\theta}\indices{_{\hat{I}}^j} &= \theta^{\hat{I}j} \tilde{\theta}\indices{_{\hat{I}}^i} .
\end{align}

Now, we make the field redefinitions
\begin{equation} \label{eq:change1}
B{''}_\alpha = Z\indices{_\alpha^\beta} B'_\beta, \qquad A^I = (Y^{-1})\indices{^I_J} A'^J, \qquad \tilde{A}_I = Y\indices{^J_I} \tilde{A}'_J .
\end{equation}
In those variables, the Lagrangian takes the same form \eqref{lagtwoforms} but with  primed quantities everywhere. The new matrices $\mathcal{I}'$ and $\mathcal{R}'$ are given by
\begin{equation}
\mathcal{I}' = Y^{-T} \mathcal{I} Y^{-1}, \qquad
\mathcal{R}' = Y^{-T} \mathcal{R} Y^{-1}.
\end{equation}
These field  redefinitions are local, so any local function of the old set of variables is also a local function of the new set of variables. 

Using equation \eqref{thetasplit}, it can be seen that the magnetic vector fields $\tilde{A}'_{\hat{U}}$ and the two-forms $B'_m$ do not appear in the Lagrangian\footnote{This means that their gauge symmetries are in fact pure shift symmetries. In particular, a complete description of the gauge symmetries of the  two-forms $B'_m$ is actually given by $\delta B'_m = \epsilon'_m$ rather than $\delta B'_m = d \Xi'_m$, implying that the above set of gauge transformations is not complete.  This is correctly taken into account in the BRST discussion of the next subsection where the BRST transformation of $B'_m$ is $s B'_m = C'_m$ with undifferentiated ghost $C'_m$.  Similar features (shift symmetries for some of the $2$-forms) hold when $g$ is turned on.}  .

Let us also redefine the gauge parameters as $\Xi'_\alpha = (Z^{-1})\indices{_\alpha^\beta} \Xi{''}_\beta$, $\lambda'^I = Y\indices{^I_J} \lambda^J$ and $\tilde{\lambda}'_I = (Y^{-1})\indices{^J_I} \tilde{\lambda}_J$. Ignoring temporarily $A'^{\hat{U}}$,  the gauge variations of $A'^{\hat{I}}$, $\tilde{A}'_{\hat{I}}$ and $B'_i$ are then
\begin{equation}
\delta A'^{\hat{I}}_\mu = \partial_\mu \lambda'^{\hat{I}} - \frac{1}{2} \theta^{\hat{I}i} \Xi'_{i\mu}, \quad \delta \tilde{A}^\prime_{\hat{I}\mu} = \partial_\mu \tilde{\lambda}'_{\hat{I}} + \frac{1}{2} \tilde{\theta}\indices{_{\hat{I}}^i} \Xi'_{i\mu}, \quad
\delta B'_{i\mu\nu} = 2 \partial_{[\mu}\Xi'_{i \nu]} .
\end{equation}
 They suggest the further changes of variables
\begin{align} \label{eq:change2}
\bar{A}^i_{\mu} &= \theta^{\hat{I}i} \tilde{A}'_{\hat{I}\mu} + \tilde{\theta}\indices{_{\hat{I}}^i} A'^{\hat{I}}_{\mu} \\
\Delta_{i\mu\nu} &= B'_{i\mu\nu} + 2 (\theta^{-1})_{i\hat{I}} F'^{\hat{I}}_{\mu\nu} ,
\end{align}
which we complete in the $A$-sector by taking other independent linear combinations of the $A'^{\hat{I}}$, $\tilde{A}'_{\hat{I}}$, which can be taken to be for instance the $A'^{\hat{I}}_\mu$.   The change of variables is again such that any local function of the old set of variables is also a local function of the new set of variables.  Using the constraint $\theta^{\hat{I}i}\tilde{\theta}\indices{_{\hat{I}}^j} = \theta^{\hat{I}j}\tilde{\theta}\indices{_{\hat{I}}^i}$, one finds that the variation of the new variables simplifies to
\begin{equation}
\delta \bar{A}^i_\mu = \partial_\mu \eta^i, \quad \delta \Delta_{i\mu\nu} = 0, \quad \delta A'^{\hat{I}}_\mu = \epsilon^{\hat{I}}_\mu
\end{equation}
with $\eta^i = \theta^{\hat{I}i} \tilde{\lambda}'_{\hat{I}} + \tilde{\theta}\indices{_{\hat{I}}^i} \lambda'^{\hat{I}}$ and $\epsilon^{\hat{I}}_\mu = \partial_\mu \lambda'^{\hat{I}} - \frac{1}{2} \theta^{\hat{I}i} \Xi'_{i\mu}$. We have used a different symbol $\Delta_{i\mu\nu}$ to emphasize that it does not transform anymore as a generalized gauge potential. Because $\theta^{\hat{I}i}$ is invertible, the gauge parameters  $\eta^i$ and $ \epsilon^{\hat{I}}_\mu$ provide an equivalent description of the gauge symmetries, but contrary to that given by $\lambda'^{\hat{I}}$, $\tilde{\lambda}'_{\hat{I}}$ and $\Xi'_{i\mu}$, it is an irreducible one.

Written in those variables, the Lagrangian only depends on $n_v$ vector fields $A'^{\hat{U}}$ and $\bar{A}^i$ and on $r$ two-forms $\Delta_i$. The variables $A'^{\hat{I}}_\mu$ drop out in agreement with the shift symmetry $\delta A'^{\hat{I}}_\mu = \epsilon^{\hat{I}}_\mu$.

 The Lagrangian is explicitly
\begin{equation} \label{eq:lbarB}
\mathcal{L} = - \frac{1}{4} \mathcal{I}'_{IJ}(\phi) \bar{H}^I_{\mu\nu} \bar{H}^{J\mu\nu} + \frac{1}{8} \mathcal{R}'_{IJ}(\phi)\, \varepsilon^{\mu\nu\rho\sigma} \bar{H}^I_{\mu\nu} \bar{H}^J_{\rho\sigma} - \frac{1}{8} \varepsilon^{\mu\nu\rho\sigma} \Delta_{i \mu\nu} \left( \bar{F}^i_{\rho\sigma} - \frac{1}{4} \tilde{\theta}\indices{_{\hat{I}}^i} \theta^{\hat{I}j} \Delta_{j \rho\sigma}\right),
\end{equation}
where the $\bar{H}^I$ are
\begin{equation}
\bar{H}^{\hat{I}} = \frac{1}{2} \theta^{\hat{I}i} \Delta_i, \quad \bar{H}^{\hat{U}} = F'^{\hat{U}} .
\end{equation}
The gauge variations of the fields are
\begin{equation}
\delta A'^{\hat{U}}_\mu = \partial_\mu \lambda'^{\hat{U}}, \quad \delta \bar{A}^i_\mu = \partial_\mu \eta^i, \quad \delta \Delta_{i\mu\nu} = 0.
\end{equation}
The two-forms $\Delta_i$ are auxiliary fields and can be eliminated from the Lagrangian \eqref{eq:lbarB}, yielding a Lagrangian of the form \eqref{eq:lprime} in a definite symplectic frame. This is because the relevant quadratic form is invertible (see \cite{deWit:2005ub}, section 5.1, where the final Lagrangian may also be found).

One can get this Lagrangian more directly as follows.  After the auxiliary fields are eliminated, the variables that remain are the scalar fields and the $n_v$ vector fields $A'^{\hat{U}}$ and $\bar{A}^i$.  We thus see that the embedding tensor determines a symplectic frame. The matrix $E \in Sp(2n_v, \mathbb{R})$ defining the symplectic frame can be viewed in this approach as the function $E(\Theta)$ of the embedding tensor obtained through the above successive steps that lead to the final Lagrangian where only half of the potentials remain. Of course, there are ambiguities in the derivation of $E$ from the embedding tensor, since choices were involved at various stages in the construction.  One gets $E$ up to a transformation of the stability subgroup of the Lagrangian subspace of the $n_v$ electric potentials  $A'^{\hat{U}}$ and $\bar{A}^i$.

More explicitly, one gets the matrix $E(\Theta)$ from the above construction as follows: the change of variables \eqref{eq:change1} can be written as
\begin{equation} \label{E1}
\begin{pmatrix}
A^I \\ \tilde{A}_I
\end{pmatrix} = \begin{pmatrix}
(Y^{-1})\indices{^I_J} & 0 \\ 0 & Y\indices{^J_I}
\end{pmatrix} \begin{pmatrix}
A'^J \\ \tilde{A}'_J
\end{pmatrix}
\end{equation}
and the full change \eqref{eq:change2} (including the trivial redefinitions of the other vector fields) as
\begin{equation} \label{E2}
\begin{pmatrix}
A'^{\hat{I}} \\ A'^{\hat{U}} \\\tilde{A}'_{\hat{I}} \\ \tilde{A}'_{\hat{U}}
\end{pmatrix} =
\begin{pmatrix}
0 & 0 & \delta^{\hat{I}}_{\hat{J}} & 0 \\
0 & \delta^{\hat{U}}_{\hat{V}} & 0 & 0 \\
\delta^{\hat{J}}_{\hat{I}} & 0 & - (\theta^{-1})_{i \hat{I}} \tilde{\theta}\indices{_{\hat{J}}^i} & 0 \\
0 & 0 & 0 & \delta^{\hat{V}}_{\hat{U}}
\end{pmatrix}
\begin{pmatrix}
\bar{A}_{\hat{J}} \\ \bar{A}^{\hat{V}} \\ \bar{A}^{\hat{J}} \\ \bar{A}_{\hat{V}}
\end{pmatrix},
\end{equation}
where we defined
\begin{equation}
\bar{A}_{\hat{I}} = (\theta^{-1})_{i \hat{I}} \bar{A}^i
\end{equation}
with respect to \eqref{eq:change2} in order to have the same kind of indices. The matrix $E(\Theta)$ is therefore simply given by
\begin{equation}
E(\Theta) = E_1 E_2,
\end{equation}
where $E_1$ and $E_2$ are the matrices appearing in equations \eqref{E1} and \eqref{E2} respectively. Using the property
\begin{equation}
(\theta^{-1})_{i[\hat{I}} \tilde{\theta}\indices{_{\hat{J}]}^i} = 0,
\end{equation}
which follows from the last of \eqref{thetasplit} upon contractions with $\theta^{-1}$, one can show that $E(\Theta)$ defined in this way is indeed a symplectic matrix. Once $E(\Theta)$ is known, the final Lagrangian in this symplectic frame (i.e., the Lagrangian that follows from the elimination of $\Delta_i$ in \eqref{eq:lbarB}) is then simply the one of section \ref{sec:symplecticchoice},  that is,  the Lagrangian \eqref{eq:lprime} with $\mathcal{I}'$ and $\mathcal{R}'$ determined from \eqref{eq:OMdefBis} where $\mathcal{M}'$ is given by 
\begin{equation}\label{eq:MprimeTheta}
\mathcal{M}' = E(\Theta)^T \mathcal{M} E(\Theta) .
\end{equation}
(see \eqref{eq:Mprime}). In general, the matrix $E$ will have an upper triangular part, and hence, will not belong to the stability subgroup of the original electric frame.

We close this subsection by noting that, as we have seen,  it is the rank of the magnetic part of the embedding tensor that controls how many original ``magnetic" fields become ``electric".  In particular, if the rank is zero, all original electric fields remain electric.  While if the rank is maximum, all original magnetic fields become electric.

\subsection{BRST cohomology}
The space of equivalence classes of non-trivial, consistent deformations of a local action is isomorphic to the BRST cohomology in field-antifield space \cite{Batalin:1981jr} at ghost number zero \cite{Barnich:1993vg}.  The cohomology  in the space of local functionals, relevant here, is denoted by $H^0(s \vert d)$.  The question, then, is to determine how the variables appearing in addition to the standard variables of the Lagrangian \eqref{eq:lprime}  could modify $H^0(s \vert d)$.

It follows from the previous subsection that these extra variables  are of two types:
\begin{itemize}
\item either they are of pure gauge  type, dropping out from the Lagrangian  -- or what is the same,   invariant under arbitrary shifts;
\item or they are auxiliary fields appearing quadratically and undifferentiated in the Lagrangian, consequently they can be eliminated algebraically through their own equations of motion.  \end{itemize}
We collectively denote by $W^\alpha$ the fields of pure gauge type.  So the $W^\alpha$'s stand for the $n_v$ vector potentials $\tilde{A}'_{\hat{U}}$, $A'^{\hat{I}}$ and the (dim$\,G-r$) two-forms $B'_m$.  The auxiliary fields are the $r$ two-forms $\Delta_i$.    For convenience, we absorb in (\ref{eq:lbarB}) the linear term in the auxiliary fields by a redefinition $\Delta_i \rightarrow \Delta_i' = \Delta_i + b_i$ where the term $b_i$ is $\Delta_i$-independent.  Once this is done, the dependence of the Lagrangian on the auxiliary fields simply reads $\frac12 \kappa^{ij} \Delta_i' \Delta_j'$ with an invertible quadratic form $\kappa^{ij}$.

The BRST differential acting in the sector of the first type of variables read
\be
s W^\alpha = C^\alpha \, ,\; \; \;   s C^\alpha = 0 \, ,\; \; \; s C^*_\alpha = W^*_\alpha \, ,\; \; \; s W^*_\alpha =0
\ee
where $C^\alpha$ are the ghosts of the shift symmetry and $W^*_\alpha$, $C^*_\alpha$ the corresponding antifields, so that $(C^\alpha,  W^\alpha)$ and $(W^*_\alpha, C^*_\alpha)$  form ``contractible pairs" \cite{Henneaux:1992ig}.  Similarly, the BRST differential acting in the sector of the second type of variables read
\be
s \Delta_i' = 0, \; \; \; s \Delta^{'*i} = \kappa^{ij} \Delta_j'
\ee
where $\Delta^{'*i}$ are the antifields conjugate to $\Delta_i'$, showing that the auxiliary fields and their antifields form also contractible pairs since $\kappa^{ij}$ is non degenerate.

Now, the above BRST transformations involve no spacetime derivatives so that one can construct a ``contracting homotopy" \cite{Henneaux:1992ig} in the sector of the extra variables $(W^\alpha,  \Delta_i')$, their ghosts and their antifields that commutes with the derivative operator $\partial_\mu$.  The algebraic setting
is in fact the same as for the variables of the ``non minimal sector" of \cite{Batalin:1981jr}.  This implies that the extra variables neither contribute to $H^k(s)$ nor to $H^k(s \vert d)$  \cite{Barnich:1994db}.

To be more specific, let us focus on the contractible pair $(C^\alpha, W^\alpha)$.  The analysis proceeds in exactly the same way for the other pairs.  One can write the BRST differential  in that sector as
\be
s = \sum_{\{ \mu \}} \partial_{\mu_1 \cdots \mu_s} C^\alpha \frac{\partial}{\partial _{\mu_1 \cdots \mu_s} W^\alpha}
\ee
where the sum is over all derivatives of the fields.  Define the ``contracting homotopy" $\rho$ as
\be
\rho = \sum_{\{ \mu \}} \partial_{\mu_1 \cdots \mu_s} W^\alpha \frac{\partial}{\partial _{\mu_1 \cdots \mu_s} C^\alpha}  .
\ee
One has by construction 
\be
[ \rho, \partial_\mu] = 0,
\ee
just as 
\be
[ s, \partial_\mu] = 0.
\ee
This is equivalent to $\rho d + d \rho = 0$.  Furthermore, the counting operator $N$ defined by
\be
N = s \rho + \rho s
\ee
is explicitly given by
\be
N = \sum_{\{ \mu \}} \partial_{\mu_1 \cdots \mu_s} C^\alpha \frac{\partial}{\partial _{\mu_1 \cdots \mu_s} C^\alpha} + \sum_{\{ \mu \}} \partial_{\mu_1 \cdots \mu_s} W^\alpha \frac{\partial}{\partial _{\mu_1 \cdots \mu_s} W^\alpha}
\ee
and commutes with the BRST differential,
\be
[N, s] = 0.
\ee
The operator $N$ gives the homogeneity degree in $W^\alpha$, $C^\alpha$ and their derivatives.  So,  the polynomial $a$ is such  $Na = k a $ with $k$ a non negative integer if and only if  it is of degree $k$ in $W^\alpha$, $C^\alpha$ and their derivatives (by Euler theorem for homogeneous functions). 

Because $N$ commutes with $s$, we can analyze the cocycle condition
\be
s a + d b = 0 \label{eq:cocycle}
\ee
at definite polynomial degree, i.e., assume that $N a = k a$, $Nb = kb$ where $k$ is a non-negative integer.  Our goal is to prove that the solutions of (\ref{eq:cocycle}) are trivial when $k \not=0$, i.e., of the form $a = se + df$, so that one can find, in any cohomological class of $H(s \vert d)$, a representative that does not depend on $W^\alpha$ or $C^\alpha$ -- that is, $W^\alpha$ and $C^\alpha$ ``drop from the cohomology".

To that end, we start from $ a = N \left(\frac{a}{k} \right)$ ($k \not=0$), which we rewrite as
$a = s \rho \left(\frac{a}{k} \right)+ \rho s  \left(\frac{a}{k} \right)$ using the definition of $N$.  The first term is equal to $s \left(\rho \left(\frac{a}{k} \right)\right)$ and hence is BRST-exact.   The second term is equal to $\rho \left(-d \left(\frac{b}{k} \right)\right)$ (using (\ref{eq:cocycle})), which is the same as $d \left(\rho \left(\frac{b}{k} \right)\right)$ since $\rho$ and $d$ anticommute.  So it is $d$-exact. We thus have shown that 
\be
a = se + df
\ee
with $e = \rho \left(\frac{a}{k} \right)$ and $f = \rho \left(\frac{b}{k} \right)$.   This is what we wanted to prove.

 We can thus conclude that the local deformations ($H^0(s \vert d)$) -- and in fact also $H^k(s \vert d)$ and in particular the candidate anomalies ($H^1(s \vert d)$) --  are the same whether or not one includes the extra fields $(W^\alpha, \bar{B}_i')$.  This is what we wanted to prove: the BRST cohomologies computed from the Lagrangians \eqref{lagtwoforms} and \eqref{eq:lprime} are the same.

We note finally that models with similar structure have been studied in \cite{Boulanger:2008nd}.

\section{Conclusions and Comments}\label{sec:conclusions}

The question of finding all consistent local deformations of supergravities is an important one.  In particular, the deep role played by electric-magnetic duality in that context  must be taken into account.

This paper, where we considered only the non-gravitational bosonic sector of supergravities to focus on the role of electric-magnetic duality, motivates the systematic analysis of the local deformations of the standard Lagrangian   \eqref{eq:lprime} with only $n_v$ vector fields, in which the symplectic frame is kept arbitrary.  By doing so, one does not miss deformations that would be available in the embedding tensor formalism (with arbitrary embedding tensor), since we have proved that both formulations admit exactly the same local BRST cohomologies, at ghost number zero (consistent deformations) but in fact also at all ghost numbers.   

We have furthermore shown that the first-order manifestly duality invariant action does not admit local deformations of the Yang-Mills type.  One must turn to the second-order action \eqref{eq:lprime} to have access to these local deformations.

In separate work \cite{preparation}, we systematically compute the local BRST cohomology of the standard Lagrangian   \eqref{eq:lprime} by methods covering any choice of symplectic frame and without making a priori assumption on the type of deformation.


\subsection*{Acknowledgements}

We are grateful to Glenn Barnich and Nicolas Boulanger for useful discussions. MH thanks the Laboratoire de Physique Th\'eorique de l'Ecole Normale Sup\'erieure for kind hospitality. BJ thanks ULB for its generous hospitality.   This work was partially supported by the ERC Advanced Grant ``High-Spin-Grav", by FNRS-Belgium (convention FRFC PDR T.1025.14 and  convention IISN 4.4503.15) and by the ``Communaut\'e Fran\c{c}aise de Belgique" through the ARC program. VL is a Research Fellow at the Belgian F.R.S.-FNRS.





\end{document}